\documentclass[10pt, conference]{IEEEtran}

\usepackage{cite}

\usepackage[pdftex]{graphicx}

\usepackage[caption=false,font=footnotesize]{subfig}

\usepackage{amsmath}
\usepackage{amsthm}
\usepackage{amssymb}
\usepackage{bm}

\usepackage{mathtools}
\usepackage{siunitx}
\usepackage{makecell}
\usepackage{url}

\usepackage{algpseudocode}

\usepackage[cmintegrals]{newtxmath}

\usepackage{lipsum}

\usepackage{siunitx}

\usepackage{soul,color}

\usepackage{mathrsfs}

\usepackage{array}

\usepackage[inline]{enumitem}

\usepackage{hyperref}

\DeclareMathAlphabet{\mathcal}{OMS}{cmsy}{m}{n}

\usepackage{multirow}
\usepackage{array}
\newcolumntype{L}[1]{>{\raggedright\let\newline\\\arraybackslash\hspace{0pt}}m{#1}}
\newcolumntype{C}[1]{>{\centering\let\newline\\\arraybackslash\hspace{0pt}}m{#1}}
\newcolumntype{R}[1]{>{\raggedleft\let\newline\\\arraybackslash\hspace{0pt}}m{#1}}

\usepackage{todonotes}
\definecolor{morange}{rgb}{0.5,0.2,0.1}
\definecolor{mblue}{rgb}{0,0.1,1.0}
\definecolor{mred}{rgb}{1,0,0}
\definecolor{mgreen}{rgb}{0.2,0.4,0}
\definecolor{navyblue}{rgb}{0.2, 0.2, 0.8}
\definecolor{mordantred19}{rgb}{0.68, 0.05, 0.0}
\definecolor{napiergreen}{rgb}{0.16, 0.5, 0.0}

\newtheoremstyle{}{}{}{\itshape}{}{\bfseries}{.}{ }{}
\newtheorem{thm}{Theorem}

\newcommand{\tmatrix}[4]{\mathbf{#1}^{\left(#2\right)#4}_{\text{#3}}}
\newcommand{\tmatrixNoPar}[4]{\mathbf{#1}^{#2#4}_{{#3}}}

\begin{document}
%

\title{Dynamic RF Beam Codebook Reduction for \\ Cost-Efficient mmWave Full-Duplex Systems}


\author{\IEEEauthorblockN{Gee Yong Suk$^1$, Jongwoo Kwak$^1$, Jaeyoung Choi$^2$, and
Chan-Byoung Chae$^1$}
\IEEEauthorblockA{$^1$School of Integrated Technology, Yonsei University, Korea \\ $^2$Samsung
	Electronics Co., Ltd., Korea \\ Email:$^1$\{gysuk, kjw8216, cbchae\}@yonsei.ac.kr, $^2$jyoung2.choi@samsung.com} 
}

\maketitle

\begin{abstract}
The recent attempts to realize full-duplex (FD) communications in millimeter wave (mmWave) systems have garnered a significant amount of interest for its potential. In this paper, we present a cost-efficient design of mmWave FD systems, where the system dynamically reduces the RF beam codebook in a computationally efficient manner, so that it is comprised of the RF beams that will prevent the Rx receive chain from saturating due to the self-interference (SI). The analog beamformer will suppress the SI to the level that the residual SI can be completely removed with digital SI cancellation, allowing the digital beamformer to concentrate on the desired channel, free of the SI. To reduce the computation required for the proposed method, we propose two sufficient conditions that prevent the Rx side saturations, which are practically tight. Through performance evaluations conducted in realistically modeled mmWave FD scenarios, we demonstrate that the proposed design achieves comparable performance with the ideal FD and other benchmarks with significantly lower costs.
\end{abstract}

\section{Introduction}
 
The large spectrum available at millimeter-wave (mmWave) frequencies offers a potential boost in transmission speed. 
Hybrid beamforming has been widely adopted as a typical architecture for the mmWave systems~\cite{Samsung2014mmWave}, where the hybrid of analog and digital beamforming balances the trade-off of multiplexing gain and hardware (H/W) cost. It was shown that hybrid beamforming can achieve a performance near that of full-digital beamforming with reasonable hardware cost by exploiting the spatially sparse channel characteristic of the mmWave system~\cite{RHeath2014spatialySparse}.

Nevertheless, these studies have been confined by the half-duplex (HD) framework, whereas the ever-increasing demand for a high data rate motivates us to consider doubling the system capacity with full-duplex (FD). The FD is one of the promising technologies for future wireless communications, which maximizes the spectral efficiency by allowing transmission and reception to occur simultaneously at the same frequency band. Furthermore, aside from doubling the spectral efficiency, the FD also has the potential to increase the flexibility in the design of wireless protocols~\cite{fodor2021guest}.

The critical technical challenge regarding the FD is the mitigation of the self-interference (SI), an undesired interfering signal at the FD-node, transmitted by itself. Due to the proximity of the interference source compared to that of the desired signal's source, the SI is a powerful interference signal that can cause intolerable performance loss. 
Conventional FD radios have extensively investigated the self-interference cancellation (SIC) methods of which a transceiver, with the knowledge of its own transmit signal, reconstructs the SI and subtracts it at the receiver. The SIC may be done at the passband via analog circuit (active analog SIC) and/or at the baseband (digital SIC)~\cite{ Sachin2013Full,kwack,prototype, kim2022performance}, aiming to suppress the SI down to the noise floor. 

Realizing the FD communications in the mmWave system introduced new approaches due to the unique characteristics of the mmWave systems~\cite{ipr2021magazine,ZXiao2017FDmmWave}. Previous works have focused on the beamforming-based FD designs, which exploit numerous antennas herein~\cite{Hanzo2019TVT,ipr2020bfc,ipr2020freq}. These works have investigated the design of analog/digital beamformers that suppresses the SI at the propagation domain, under the assumption that the full channel state information (CSI) of the desired channel is given. Furthermore, designs in \cite{ipr2020freq,ipr2020bfc} took the advantage of an increased number of RF chains in order to provide the sufficient dimensionality of the SI channel's null space.
However, these solutions are rather infeasible because practical systems generally rely on beam training with finite-resolution RF beam codebooks, and the number of RF chains would typically match the number of transmitting and receiving data streams. 

The joint design of the SIC and beamforming for the mmWave FD systems appears favorable, as the responsibility of mitigating the SI is shared across the SIC modules and hybrid beamformers~\cite{ipr2021magazine}. By offloading the burden of mitigating the SI, beamformers can better  concentrate on the desired links.
The authors in \cite{ipr2021TWC} proposed the joint design, where the hybrid beamformer was acquired by the repitition of solving a convex optimization problem under the constraints that prevent the receiver side saturations. However, the design in \cite{ipr2021TWC} is costly, raising concerns for the feasibility of real-time implementations.
Besides, considering that the typical orthogonal frequency division multiplexing (OFDM) will require the computational complexity proportional to the number of subcarriers, it is imperative to devise a cost-efficient design for the mmWave FD systems.

\begin{figure*}[t]
	\centerline{\resizebox{2\columnwidth}{!}{\includegraphics{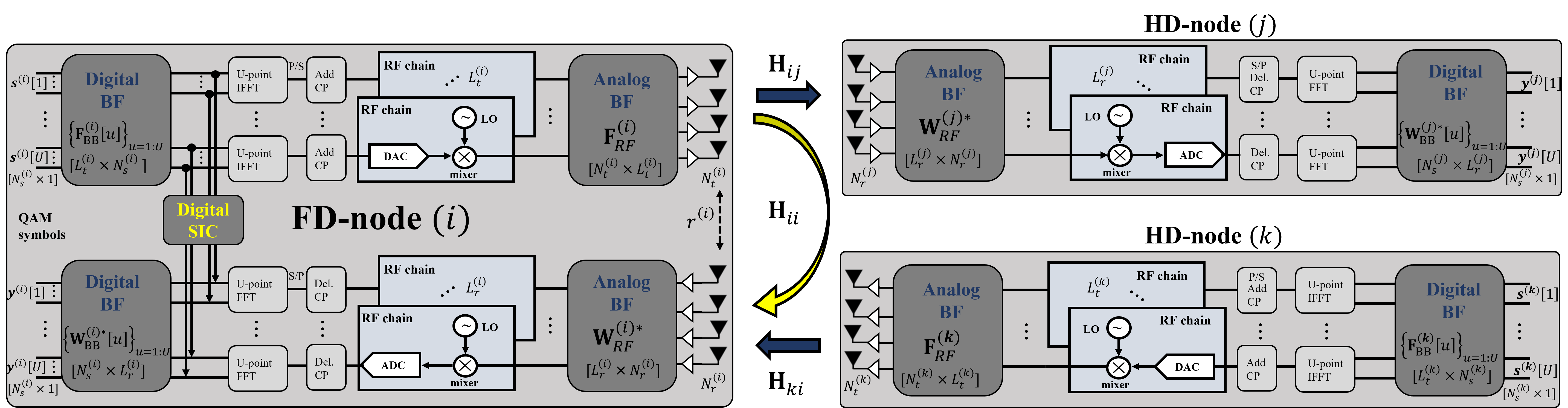}}}
	\caption{A block diagram of the mmWave FD OFDM system with two HD-nodes and one FD-node, where the FD-node employs a joint design of analog/digital beamforming (BF) and the digital SIC.}	
	\label{Fig.sysTemModel}
\end{figure*}

In this paper, we propose a cost-efficient design of the mmWave FD systems with the dyanmic RF beam codebook reduction to lower the required number of beam measurements with the minimum computations. Specifically, a reduced-size RF beam codebook is acquired from a set of RF beam combinations that solely prevent the Rx side saturations. Then, the joint operation of the digital SIC allows the digital beamformer design to focus on the desired channel, free of the SI. We propose two sufficient conditions for preventing the Rx side saturations with analog beamformers that involve significantly reduced computational overhead, which are yet practically tight. We evaluate the system performance in practical mmWave FD scenarios and demonstrate that the proposed design offers comparable performance with the ideal FD and other leading benchmarks with significantly lower cost.  

\section{System Model}
\label{sec.2SIC_Anal}

This work considers a mmWave FD system that consists of FD-node $i$ and  HD-nodes $j$ and $k$, as illustrated in Fig.~\ref{Fig.sysTemModel}. Node $i$ simultaneously transmits data streams to node~$j$, while receiving data streams from node $k$ in the same frequency band. Node $i$ performs the SI mitigation via the joint cooperation of beamforming and SIC.
The nodes considered in the system deploy hybrid beamforming in a fully-connected manner, where each antenna element is connected to each RF chain through a network of phase shifters. Also, we assume that node $i$ uses separate arrays for its transmission and reception, separated by distance $r^{\left(i\right)}$, which is one of the factors that governs the amount of passive SI suppression.

To cover the wide spectrum of the mmWave channel, we consider a wideband system with the OFDM, along with $U$ subcarriers. For the node index $(m,n)\in \{(i,j),(k,i)\}$, let  $N_t^{(m)},N_r^{(n)}$,  $L_t^{(m)},L_r^{(n)}$, and $N_s^{\left(mn\right)}$ denote the number of transmit and receive antennas, the number of the Tx and Rx RF chains, and the number of symbol streams that are transmitted by $m$ and received by $n$, respectively.
The symbol streams transmitted by the $u$-th subcarrier of node $m$ are denoted by $\textbf{s}^{(m)}[u]\in\mathbb{C}^{N_s^{\left(mn\right)}\times 1}$, and the symbol streams received by the $u$-th subcarrier of $n$ are denoted by $\textbf{y}^{(n)}[u]\in\mathbb{C}^{N_s^{\left(mn\right)}\times 1}$, respectively, where $u\in\{1,2,\dots,U\}$. Accordingly, we denote the digital and analog precoding matrices by $F_{\text{BB}}^{(m)}[u]\in\mathbb{C}^{L_t^{(m)}\times N_s^{(m)}}$ and $F_{\text{RF}}^{(m)}\in\mathbb{C}^{N_t^{(m)}\times L_t^{(m)}}$, the digital and analog combining matrices by $W_{\text{BB}}^{(n)}[u]\in\mathbb{C}^{L_r^{(n)}\times N_s^{(n)}}$ and $W_{\text{RF}}^{(n)}\in\mathbb{C}^{N_r^{(n)}\times L_r^{(n)}}$, respectively. Unlike the digital precoder/combiners, the analog precoder/combiners are frequency-flat across the entire $U$ subcarriers. 
The symbol vector, $\textbf{s}^{(m)}[u]\sim\mathcal{N}_{\mathbb{C}}(\textbf{0},\frac{1}{N_s^{\left(mn\right)}}\cdot\tmatrixNoPar{I}{}{}{}{})$, assumes Gaussian signaling and equal power distribution among the subcarriers. Let $\textbf{n}^{(n)}[u]\sim\mathcal{N}_{\mathbb{C}}(\textbf{0},\sigma_n^2\cdot\tmatrixNoPar{I}{}{}{}{})$ be the additive noise vector at the array of receiver $n$ per subcarrier with noise variance $\sigma_n$. Let $\text{SNR}_{mn}=P_\text{tx}^{\left(m\right)}\cdot G_{mn}^2/\sigma_n^2$ be the signal-to-noise ratio (SNR) without any beamforming gains from node $m$ to $n$, where $P_\text{tx}^{\left(m\right)}$ is the transmit power at $m$ and $G_{mn}^2$ is the large-scale power gain of the propagation from $m$ to $n$.

For practical considerations, we select the unit-norm columns of $\tmatrix{F}{m}{RF}{}$ and $\tmatrix{W}{n}{RF}{}$ from the predefined RF beam codebooks that account for the H/W constraints such as phase shifter resolution and lack of amplitude control. We define
\begin{equation}
	\big[\tmatrix{F}{m}{RF}{}\big]_{:,\ell} \in \tmatrix{\mathcal{F}}{m}{RF}{}, \quad\ell=1,2,\dots,L_t^{\left(m\right)}, 
\end{equation}
\begin{equation}
	\big[\tmatrix{W}{n}{RF}{}\big]_{:,\ell} \in \tmatrix{\mathcal{W}}{n}{RF}{},\quad\ell=1,2,\dots,L_r^{\left(n\right)}, 
\end{equation}
where $\tmatrix{\mathcal{F}}{m}{RF}{}$ and $\tmatrix{\mathcal{W}}{n}{RF}{}$ are the analog precoding and combining codebooks, respectively. Also, we define $C_t^{\left(m\right)}$ and $C_r^{\left(n\right)}$ as the number of candidate beams in $\tmatrix{\mathcal{F}}{m}{RF}{}$ and $\tmatrix{\mathcal{W}}{n}{RF}{}$, respectively. Then, the transmitter's total power constraint is enforced by
\begin{equation}
\label{eq:powerConst}
	\left\lVert\tmatrix{F}{m}{BB}{}[u]\right\rVert^2 \leq N_s^{\left(mn\right)}.
\end{equation}

For the desired channels $\tmatrixNoPar{H}{}{ij}{}$ and $\tmatrixNoPar{H}{}{ki}{}$, we use the far-field channel model based on the Saleh-Valenzuela model~\cite{RHeath2014spatialySparse}, where the multiple-input multiple-output (MIMO) channel matrix between the transmit array and receive array at tap $d$ is
\begin{equation}
	\label{eq:chn_farfield}
	\tmatrixNoPar{H}{\text{FF}}{d}{}[d]= \sum_{\ell=1}^{N_\text{ray}}\alpha_\ell\delta(dT_s-\tau_\ell)\tmatrixNoPar{a}{}{\text{r}}{}(\theta^\text{AoA}_\ell, \phi^\text{AoA}_\ell)\tmatrixNoPar{a}{*}{\text{t}}{}(\theta^\text{AoD}_\ell, \phi^\text{AoD}_\ell),
\end{equation}
where $\alpha_\ell$ is the complex gain, $\tau_\ell$ is the time delay, $\tmatrixNoPar{a}{}{\text{r}}{}(\cdot)$ and $\tmatrixNoPar{a}{}{\text{t}}{}(\cdot)$ are the receive and transmit array responses of the elevation and azimuth angle-of-arrival (AoA), $\theta^\text{AoA}_\ell, \phi^\text{AoA}_\ell$, and the elevation and azimuth angle-of-departure (AoD), $\theta^\text{AoD}_\ell, \phi^\text{AoD}_\ell$, for the $\ell$-th path, respectively. The sampling rate of the channel is $1/T_s$ samples per second. 
For the SI channel $\tmatrixNoPar{H}{}{ii}{}$, we consider the following expression with the Rician factor $\kappa$ as
\begin{equation}
	\label{eq:chn_SI}
	\tmatrixNoPar{H}{}{d,ii}{}[d]= \sqrt{\frac{\kappa}{\kappa+1}}\tmatrixNoPar{H}{\text{NF}}{d,ii}{}+\sqrt{\frac{1}{\kappa+1}}\tmatrixNoPar{H}{\text{FF}}{d,ii}{}[d].
\end{equation}
where the ($p,q$)-th entry of the near-field channel $\tmatrixNoPar{H}{\text{NF}}{d,ii}{}$ is modeled as 
\begin{equation}
	[\tmatrixNoPar{H}{\text{NF}}{d,ii}{}]_{p,q} = \frac{\rho}{r_{p,q}}e^{-j2\pi\frac{r_{p,q}}{\lambda_c}},
\end{equation}
where $\rho$ is the normalization constant, $r_{p,q}$ is the distance between the $p$-th element of the transmit array to the $q$-th element of the receive array, $\lambda_c$ is the carrier wavelength.
Since we consider an OFDM system, we define a $U$-point discrete Fourier transform (DFT) of the channel matrices across time to get the MIMO channels across the subcarriers as
\begin{equation}
	\label{eq:chn_freq_SI}
	\tmatrixNoPar{H}{}{}{}[u]= \sum_{d=1}^{D}\tmatrixNoPar{H}{}{\text{d}}{}[d]e^{-j\frac{2\pi ud}{U}}.
\end{equation}


\section{\fontsize{11}{14}\selectfont{Proposed mmWave FD Design}}
\label{sec.4P_Proposed}

\subsection{Sufficient Conditions for Preventing Rx Saturation}
Receive chains at the FD node are especially vulnerable to saturation, due to the overwhelming SI~\cite{ipr2021TWC}. The SI can introduce nonlinearities at the low noise amplifiers (LNA) by imposing the input power level beyond its capacity, and/or degrade the quality of the desired receive signal by increasing the quantization noise at the analog-to-digital converters (ADC).
Thus, it is necessary to mitigate the portion of the SI before it reaches the receive chain. Once the system sufficiently mitigates the SI to the levels that prevent the receive saturation, and thus secure the linearities of the receive chain, the digital SIC can effectively cancel out the residual SI at the baseband~\cite{DSIC}.

The authors in \cite{ipr2021TWC} have well established the conditions to prevent the Rx saturation in a narrowband system. In this section, we expand these conditions to cover the wideband OFDM system and further formulate sufficient conditions through Theorem~\ref{theorem}, where its usage and advantage will be later discussed. Note from Fig.~\ref{Fig.sysTemModel} that the LNAs are located per-antenna and the ADCs per-RF chain. To prevent saturation at the LNAs and ADCs, the SI entering each antenna and RF chain should be suppressed below certain power thresholds. For the LNAs, achieving this in expectation over $\tmatrix{s}{i}{}{}[u]$ can be expressed as
\begin{equation}
	\label{eq:LNA_const_C1}
\begin{split}
	\sum_{u=1}^{U}\text{diag}\Big(\tmatrixNoPar{H}{}{ii}{}[u]\tmatrix{F}{i}{RF}{}\tmatrix{F}{i}{BB}{}[u]\tmatrix{F}{i}{BB}{*}[u]\tmatrix{F}{i}{RF}{*}\tmatrixNoPar{H}{}{ii}{*}[u]\Big) \\ \leq \frac{U~N_\text{s}^{\left(ij\right)} P^{max}_\text{LNA}}{P_{\text{Tx}}^{\left(i\right)} G_{ii}^2}\cdot \tmatrixNoPar{1}{}{N_{\text{r}}^{\left(i\right)}}{},
\end{split}
\end{equation}
where $P^{max}_\text{LNA}$ is the maximum power level to prevent the LNA saturation. 
For the ADCs,
\begin{equation}
	\label{eq:ADC_const_C1}
	\begin{split}
		\sum_{u=1}^{U}\text{diag}\Big(\tmatrix{W}{i}{RF}{*}\tmatrixNoPar{H}{}{ii}{}[u]\tmatrix{F}{i}{RF}{}\tmatrix{F}{i}{BB}{}[u]\tmatrix{F}{i}{BB}{*}[u]\tmatrix{F}{i}{RF}{*}\tmatrixNoPar{H}{}{ii}{*}[u]\tmatrix{W}{i}{RF}{}\Big) \\ \leq \frac{U~N_\text{s}^{\left(ij\right)} P^{max}_\text{ADC}}{P_{\text{Tx}}^{\left(i\right)} G_{ii}^2}\cdot \tmatrixNoPar{1}{}{L_{\text{r}}^{\left(i\right)}}{},
	\end{split}
\end{equation}
where $P^{max}_\text{ADC}$ is the maximum power level to prevent the ADC saturation.
Exploiting the majorization property of a Hermitian matrix, the respective sufficient condtions for \eqref{eq:LNA_const_C1} and \eqref{eq:ADC_const_C1} are given as
\begin{equation}
	\label{eq:LNA_const_C2}
	\sum_{u=1}^{U}\sigma^2_\text{max}\left(\tmatrixNoPar{H}{}{ii}{}[u]\tmatrix{F}{i}{RF}{}\tmatrix{F}{i}{BB}{}[u]\right)\leq N_\text{s}^{\left(ij\right)}\cdot\eta_\text{LNA}
\end{equation} 
and 
\begin{equation}
	\label{eq:ADC_const_C2}
	\sum_{u=1}^{U}\sigma^2_\text{max}\left(\tmatrix{W}{i}{RF}{*}\tmatrixNoPar{H}{}{ii}{}[u]\tmatrix{F}{i}{RF}{}\tmatrix{F}{i}{BB}{}[u]\right)\leq N_\text{s}^{\left(ij\right)}\cdot\eta_\text{ADC},
\end{equation}
where $\eta_\text{LNA}=\frac{U~ P^{max}_\text{LNA}}{P_{\text{Tx}}^{\left(i\right)} G_{ii}^2}$ and  $\eta_\text{ADC}=\frac{U~P^{max}_\text{ADC}}{P_{\text{Tx}}^{\left(i\right)} G_{ii}^2}$. Also, $\sigma_{\text{max}}^2(\cdot)$ denotes the maximum singular value of a given input matrix.
From $\sigma_{\text{max}}^2\left(\tmatrix{F}{i}{BB}{}[u]\right)\leq {\left\lVert\tmatrix{F}{i}{BB}{}[u]\right\rVert}^2_{\text{F}}\leq N_\text{s}^{\left(ij\right)}$, the respective sufficient condtions for \eqref{eq:LNA_const_C2} and \eqref{eq:ADC_const_C2} can be similarly derived from \cite{ipr2021TWC} as
\begin{equation}
	\label{eq:LNA_const_C3}
	\sum_{u=1}^{U}\sigma^2_\text{max}\left(\tmatrixNoPar{H}{}{ii}{}[u]\tmatrix{F}{i}{RF}{}\right)\leq \eta_\text{LNA}
\end{equation} 
and 
\begin{equation}
	\label{eq:ADC_const_C3}
	\sum_{u=1}^{U}\sigma^2_\text{max}\left(\tmatrix{W}{i}{RF}{*}\tmatrixNoPar{H}{}{ii}{}[u]\tmatrix{F}{i}{RF}{}\right)\leq \eta_\text{ADC}.
\end{equation}
\begin{table*}[t]
	\caption{Comparative analysis of required beam measurement number and computational cost for various system designs.}
	\label{table_results}
	\begin{center}
		\begin{tabular}{c|c|c|c}
			\hline
			Design methods&\makecell{Required beam measurements \\ ($@$Transmit link)}& Major computation-inducing functions & \makecell{CPU \\ run time}\\
			\hline\hline		
			Conventional~\cite{ipr2021TWC}&$64\times64~(C_r^{\left(j\right)}\times C_t^{\left(i\right)})$&Solving convex optimization problem (by CVX SeDuMi~\cite{CVXtool})&76.778~sec\\
			\hline
			
			Power reduction&$64\times64~(C_r^{\left(j\right)}\times C_t^{\left(i\right)})$&Testing saturation (w/ \eqref{eq:LNA_const_C1},\eqref{eq:ADC_const_C1})&0.058~sec\\
			\hline
			
			Proposed (w/ \eqref{eq:LNA_const_C3}, \eqref{eq:ADC_const_C3})&$64\times40.16~(C_r^{\left(j\right)}\times C_{\text{AL},t}^{\left(i\right)})$&Constructing allowlist (w/ \eqref{eq:LNA_const_C3}, \eqref{eq:ADC_const_C3})&26.471~sec\\
			\hline
			
			\textbf{Proposed (w/ \eqref{eq:LNA_const_C4}, \eqref{eq:ADC_const_C4})}&\boldmath$64\times39.33$\boldmath$~(C_r^{\left(j\right)}\times C_{\textbf{AL},t}^{\left(i\right)})$&\textbf{Constructing allowlist (w/ \eqref{eq:LNA_const_C4}, \eqref{eq:ADC_const_C4})}&\textbf{0.223~sec}\\
			\hline
		\end{tabular}
	\end{center}
\end{table*}

Finally, we define the sufficient conditions for \eqref{eq:LNA_const_C3} and \eqref{eq:ADC_const_C3} to reduce the related computational complexity that will later be discussed in detail.
\begin{thm}
	\label{theorem}
	 The following conditions, \eqref{eq:LNA_const_C4} and \eqref{eq:ADC_const_C4}, are the  
	 sufficient conditions for the per-LNA and per-ADC constraints, i.e., \eqref{eq:LNA_const_C3}, \eqref{eq:ADC_const_C3}, respectively.
\end{thm}
\begin{equation}
\label{eq:LNA_const_C4}
\sum_{\ell=1}^{L_t^{(i)}}\sum_{u=1}^{U}\left\lVert\tmatrixNoPar{H}{}{ii}{}[u]\mathbf{f}^{\left(i\right)}_{\text{RF},\ell}\right\rVert^2\leq \eta_\text{LNA},
\end{equation}
\begin{equation}
	\label{eq:ADC_const_C4}
	\sum_{\ell=1}^{L_t^{(i)}}\sum_{u=1}^{U}\left\lVert\tmatrix{W}{i}{RF}{*}\tmatrixNoPar{H}{}{ii}{}[u]\mathbf{f}^{\left(i\right)}_{\text{RF},\ell}\right\rVert^2\leq \eta_\text{ADC},
\end{equation}
where $\mathbf{f}^{\left(i\right)}_{\text{RF},\ell}$ is the $\ell$-th column vector of $\tmatrix{F}{i}{RF}{}$, i.e., $\mathbf{f}^{\left(i\right)}_{\text{RF},\ell} = \big[\tmatrix{F}{i}{RF}{}\big]_{:,\ell}$.
\begin{proof}
	Note that for a matrix $\tmatrixNoPar{A}{}{}{}\in\mathbb{C}^{N_\text{row}\times N_\text{col}}$,
	\begin{equation}
		\sigma_{\text{max}}^2(\tmatrixNoPar{A}{}{}{})\leq\sum_{n=1}^{N_\text{col}}\sigma_{\text{max}}^2\Big(\big[\tmatrixNoPar{A}{}{}{}\big]_{:,n}\Big).
	\end{equation}
	Then,  
	\begin{equation}
		\label{eq:proof}
			\begin{aligned}
					\eta_\text{LNA} &\geq \sum_{\ell=1}^{L_t^{(i)}}\sum_{u=1}^{U}\left\lVert\tmatrixNoPar{H}{}{ii}{}[u]\mathbf{f}^{\left(i\right)}_{\text{RF},\ell}\right\rVert^2
= \sum_{u=1}^{U}\sum_{\ell=1}^{L_t^{(i)}}\sigma_{\text{max}}^2(\tmatrixNoPar{H}{}{ii}{}[u]\mathbf{f}^{\left(i\right)}_{\text{RF},\ell})\\
					&\geq \sum_{u=1}^{U}\sigma_{\text{max}}^2\left(\Big[\tmatrixNoPar{H}{}{ii}{}[u]\mathbf{f}^{\left(i\right)}_{\text{RF},1}\: \tmatrixNoPar{H}{}{ii}{}[u]\mathbf{f}^{\left(i\right)}_{\text{RF},2}\: \dots \:\tmatrixNoPar{H}{}{ii}{}[u]\mathbf{f}^{\left(i\right)}_{\text{RF},L_t^{(i)}}\Big]\right)\\
					&= \sum_{u=1}^{U}\sigma_{\text{max}}^2\left(\tmatrixNoPar{H}{}{ii}{}[u]\tmatrix{F}{i}{RF}{}\right).
				\end{aligned}
		\end{equation}
	Thus,~\eqref{eq:LNA_const_C4} is a sufficient condition for \eqref{eq:LNA_const_C3}. By replacing $\tmatrixNoPar{H}{}{ii}{}[u]$ in~\eqref{eq:proof} with $\tmatrix{W}{i}{RF}{*}\tmatrixNoPar{H}{}{ii}{}[u]$, one can show that~\eqref{eq:ADC_const_C4} is a sufficient condition for \eqref{eq:ADC_const_C3}. 
\end{proof}

\subsection{Design and Operation of RF Beam Allowlists}

Our system design is motivated by \eqref{eq:LNA_const_C3} and \eqref{eq:ADC_const_C3}. These are the sufficient conditions for preventing the Rx saturation, which can be met without considering the digital precoder $\Big\{\tmatrix{F}{i}{BB}{}[u]\Big\}_{u=1:U}$, given \eqref{eq:powerConst}. This implies that there exist rooms of design to prevent the Rx saturation with only analog beamforming. Thus, our design requires the FD-node to form the analog beamformers $\tmatrix{F}{i}{RF}{}, \tmatrix{W}{i}{RF}{}$ that solely prevent the Rx saturations, and at the same time, maintain the strongest available beam alignments at its respective transmit/receive link. The Rx saturation-preventing conditions may either be \eqref{eq:LNA_const_C3} and \eqref{eq:ADC_const_C3}, or \eqref{eq:LNA_const_C4} and \eqref{eq:ADC_const_C4}. Then, the SI can be completely removed with the additional support from the digital SIC, while the digital beamformer concentrates on the desired channel, free of the SI. 

\begin{figure}[t]
	\centerline{\resizebox{0.95\columnwidth}{!}{\includegraphics{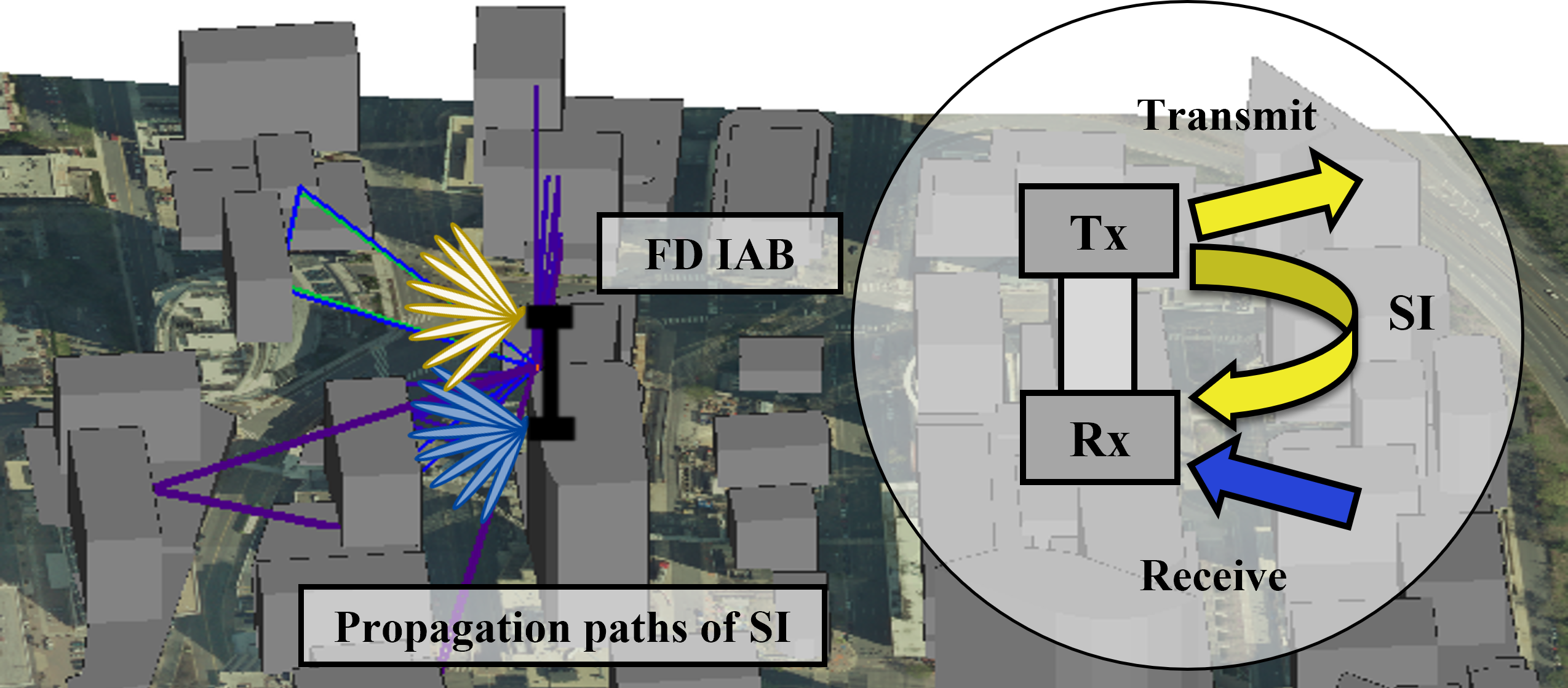}}}
	\caption{{Visual example of 3D modeling and ray-tracing algorithm for generating the SI channels of FD IAB scenarios.}}
	\label{Fig.Raytracing}	
\end{figure}

Specifically, the FD-node should periodically update its \textit{feasible beam combination set}, which is a set of beam combinations that form an analog beamformer satisfying certain Rx saturation-preventing conditions. Then, an RF beam codebook with reduced size can be formed by the set of the unique beams contained in the feasible beam combination set, which we call the \textit{allowlist}. As a toy example, consider an original RF beam codebook set comprised of 5 beams $\tmatrixNoPar{\mathcal{C}}{}{}{}=\Big\{\tmatrixNoPar{c}{}{1}{}, \tmatrixNoPar{c}{}{2}{}, \cdots, \tmatrixNoPar{c}{}{5}{}\Big\}$. Suppose that the feasible beam combination set satisfying the given Rx saturation-preventing conditions is $\tmatrixNoPar{\mathcal{P}}{}{}{}=\Big\{\left(\tmatrixNoPar{c}{}{2}{}, \tmatrixNoPar{c}{}{3}{}, \tmatrixNoPar{c}{}{4}{}\right), \left(\tmatrixNoPar{c}{}{3}{}, \tmatrixNoPar{c}{}{4}{},\tmatrixNoPar{c}{}{5}{}\right)\Big\}$, where the number of the RF chains is 3. Then, the allowlist is the reduced RF beam codebook $\mathcal{C}_{\text{AL}}=\Big\{\tmatrixNoPar{c}{}{2}{}, \tmatrixNoPar{c}{}{3}{}, \tmatrixNoPar{c}{}{4}{}, \tmatrixNoPar{c}{}{5}{}\Big\}$, necessary for the HD fashion beam sweeping measurements. Thus, with the reduced number of beam measurements based on $\mathcal{C}_{\text{AL}}$, beam combination that maximizes the beam alignment of the desired link will be selected from $\tmatrixNoPar{\mathcal{P}}{}{}{}$. The resultant RF beam combination will inherently prevent the SI from saturating the Rx components. Fig.~\ref{Fig.Raytracing} illustrates an example of differently realized allowlists.

In the actual design, $\tmatrixNoPar{\mathcal{C}}{}{}{}$ may either be $\tmatrix{\mathcal{F}}{i}{RF}{}$ or $\tmatrix{\mathcal{W}}{i}{RF}{}$.
The allowlist may be deployed at the transmit link alone or at both the transmit and receive link. For the former case, the Rx saturation-preventing conditions (\eqref{eq:LNA_const_C3} and \eqref{eq:ADC_const_C3}, or \eqref{eq:LNA_const_C4} and \eqref{eq:ADC_const_C4}) will restrain only the transmit link, while the latter will first restrain the transmit link with the LNA saturation-preventing condition (\eqref{eq:LNA_const_C3} or \eqref{eq:LNA_const_C4}), then the receive link with the ADC saturation-preventing condition (\eqref{eq:ADC_const_C3} or \eqref{eq:ADC_const_C4}) for the predetermined analog precoder $\tmatrix{F}{i}{RF}{}$. Although the latter case is expected to be more practical considering the typical load balance between the downlink and uplink in a base station scenario, we focus our analysis on the former case for brevity. 



\begin{figure}[t]
	\centerline{\resizebox{0.95\columnwidth}{!}{\includegraphics{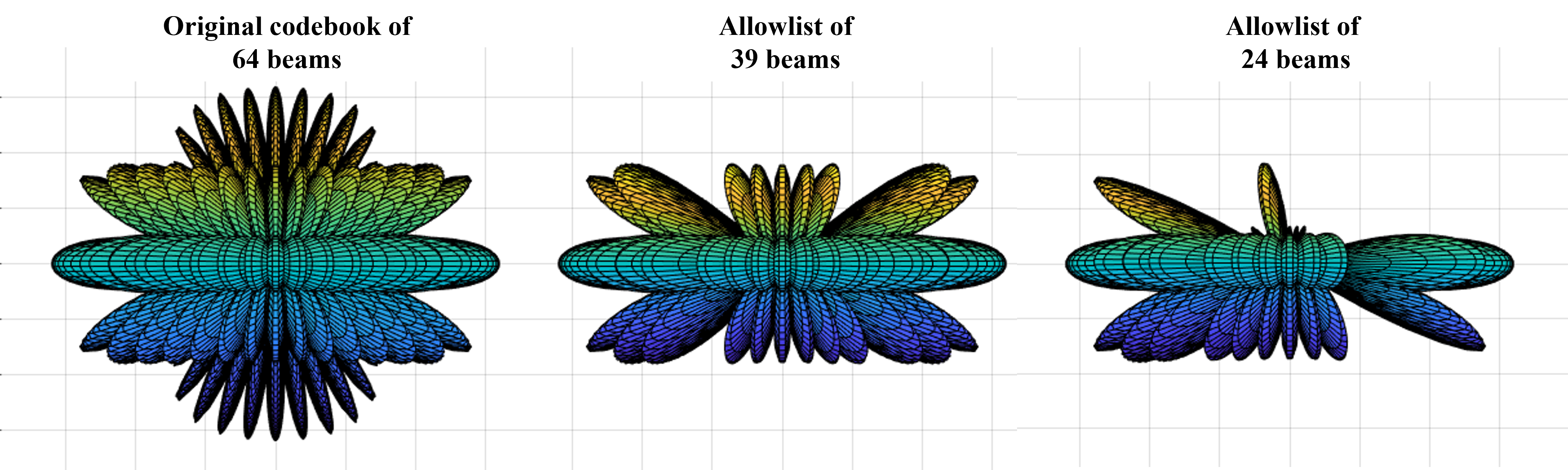}}}
	\caption{The array factors of the original $64$ beam DFT codebook and the two different realizations of allowlist with a $16\times4$ uniform planar array.}
	\label{Fig.result_ADCbit}	
\end{figure}

%
%
%

\subsection{Costs of System Design}
In this subsection, we discuss the major cost factors for implementing the proposed beamforming designs; the number of required beam measurements and the computational complexity\footnote{Computational loads for digital SIC, overheads for the estimation of the SI channel are also the cost factors that should be considered in practice. However, We omit these factors in our comparison since they are common and nearly identical for all the considered methods.}. As a benchmark, we consider (i) the expanded version of the method proposed by \cite{ipr2021TWC} for OFDM, which acquires the digital precoder per subcarrier as the solution of a convex optimization problem maximizing the transmit spectral efficiency subjected to Rx preventing conditions, \eqref{eq:LNA_const_C2} and \eqref{eq:ADC_const_C2}\footnote{The maximization problem formulated in \cite{ipr2021TWC} consists of two sequential optimization problems, where solving the inner convex problem repeats for the number of predefined \textit{candidate beam pairs}. Due to its time complexity, we only consider the case with single candidate beam pair that maximizes the beam alignment at the beam sweeping phase.}; (ii) Power reduction method, which simply reduces the transmit power to ensure that \eqref{eq:LNA_const_C1} and \eqref{eq:ADC_const_C1} are not violated, given the preselected analog beamformer that maximizes the beam alignment. 

For the proposed methods, dynamic construction of allowlist to reduce the required number of beam measurements becomes the major compuational burden. Applying \eqref{eq:LNA_const_C3} and \eqref{eq:ADC_const_C3} constraints leads to ${}_{C_t^{\left(i\right)}} C_{L_t^{\left(i\right)}}$ singular value decomposition (SVD) operations per subcarrier $\sim O$\Big($U(L_t^{\left(i\right)})^2 N_t^{\left(i\right)} {}_{C_t^{\left(i\right)}} C_{L_t^{\left(i\right)}}$\Big), which is copmutationally heavy. However, we could significantly reduce the required computations by applying \eqref{eq:LNA_const_C4} and \eqref{eq:ADC_const_C4}, leveraging their structure. Vector norm operation, which is by itself computationally lighter than the SVD, can be called for only $C_t^{\left(i\right)}$ times per subcarrier, while the condition testing is replaced by the summation of scalar norm values\footnote{Furthermore, \eqref{eq:LNA_const_C4} and \eqref{eq:ADC_const_C4} also allow us to avoid the naive approach of exhaustively computing all possible combinations using some tricks, although it is not applied in our analysis for the sake of fairness. By exploiting the reverse-lexicograhical order of the evaluated $C_t^{\left(i\right)}$ norm values, each combination can be added successively until the given constraint value is met. After this point, a chunk of combinations can be skipped, since they will also meet the given constraint value~\cite{comboGeneral}.} $\sim$${O}$\Big($U\cdot C_t^{\left(i\right)}\cdot N_t^{\left(i\right)}+{}_{C_t^{\left(i\right)}} C_{L_t^{\left(i\right)}}$\Big). 
Table~\ref{table_results} summarizes the overall comparison of the cost factors for various system designs, where the specific values are the averaged results from multiple simulations with the detailed parameter settings analogous to those described in the next section. The CPU run time is measured via Intel i7-6700K processor.

\subsection{Digital SIC and Baseband Beamforming Design}
With the designated analog precoder/combiners at the transmit/receive links $\tmatrix{F}{i}{RF}{}$, $\tmatrix{W}{j}{RF}{}$, $\tmatrix{F}{k}{RF}{}$, and $\tmatrix{W}{i}{RF}{}$, the remaining design are that of the digital SIC, the digital beamformers $\{\tmatrix{F}{i}{BB}{}[u]\}_{u=1:U}$, $\{\tmatrix{W}{j}{BB}{}[u]\}_{u=1:U}$, $\{\tmatrix{F}{k}{BB}{}[u]\}_{u=1:U}$, and $\{\tmatrix{W}{i}{BB}{}[u]\}_{u=1:U}$. First, let us define the effective SI channel as
\begin{equation}
	\label{eq:effectiveSIChn}
	\tmatrixNoPar{\tilde{H}}{}{ii}{}[u]=\tmatrix{W}{i}{RF}{*}\tmatrixNoPar{H}{}{ii}{}[u]\tmatrix{F}{i}{RF}{}.
\end{equation}
Having preserved the linearities of the receive chain, the digital SIC can effectively cancel out the residual SI at the baseband by reconstructing the residual SI with the knowledge of \eqref{eq:effectiveSIChn}~\cite{DSIC}.

Finally, owing to the analog beamformers and the digital SIC module, the design of the digital beamforming can completely focus on the effective desired channel, from node $m$ to $n$, where $(m,n)\in \{(i,j),(k,i)\}$, defined as
\begin{equation}
	\tmatrixNoPar{\tilde{H}}{}{mn}{}[u]=\tmatrix{W}{n}{RF}{*}\tmatrixNoPar{H}{}{mn}{}[u]\tmatrix{F}{m}{RF}{}.
\end{equation}
For evaluation, we considered eigenbeamforming for the digital beamformer designs, i. e., $\tmatrixNoPar{U}{}{mn}{}[u]{\boldsymbol{\Sigma}}_{mn}[u]\tmatrixNoPar{V}{}{mn}{}[u]=\text{SVD}\left(\tmatrixNoPar{\tilde{H}}{}{mn}{}[u]\right)$ and $\tmatrix{W}{n}{BB}{}[u]=[\tmatrixNoPar{U}{}{mn}{}[u]]_{1:L_r^{\left(n\right)}}$ and $\tmatrix{F}{m}{BB}{}[u]=[\tmatrixNoPar{V}{}{mn}{}[u]]_{1:L_t^{\left(m\right)}}$

\section{Performance Evaluation}
In this section, we evaluate the performance of the proposed system in a realistic integrated access and backhaul (IAB) scenario, which we envision as one of the promising applications for the mmWave FD~\cite{Suk2020FDIAB}. The IAB is a new network framework being investigated by the 3GPP, where only a few of the BSs are connected to the traditional wired infrastructures while the other BSs relay the backhaul traffic wirelessly. To precisely model the SI channel under the given network scenario, we adopt ray-tracing algorithms.
Shown in Fig.~\ref{Fig.Raytracing} is a 3D model of an urban area as a visual example. As such, we randomly pick a total of 269 points of potential FD base station positions covering urban, suburban, and rural spots in the 3D model of Texas, in the U.S. The ray-tracing software outputs, the attenuation, time delay, elevation/azimuth AoA/AoDs of maximum 15 paths, which are used to generate the MIMO SI channel using \eqref{eq:chn_farfield} and \eqref{eq:chn_SI}. The apparant characteristic of the considered SI channels is the dominance of the line-of-sight (LoS) component and the first none-line-of-sight (NLoS) componet. 

\begin{figure}[t]
	\centerline{\resizebox{0.95\columnwidth}{!}{\includegraphics{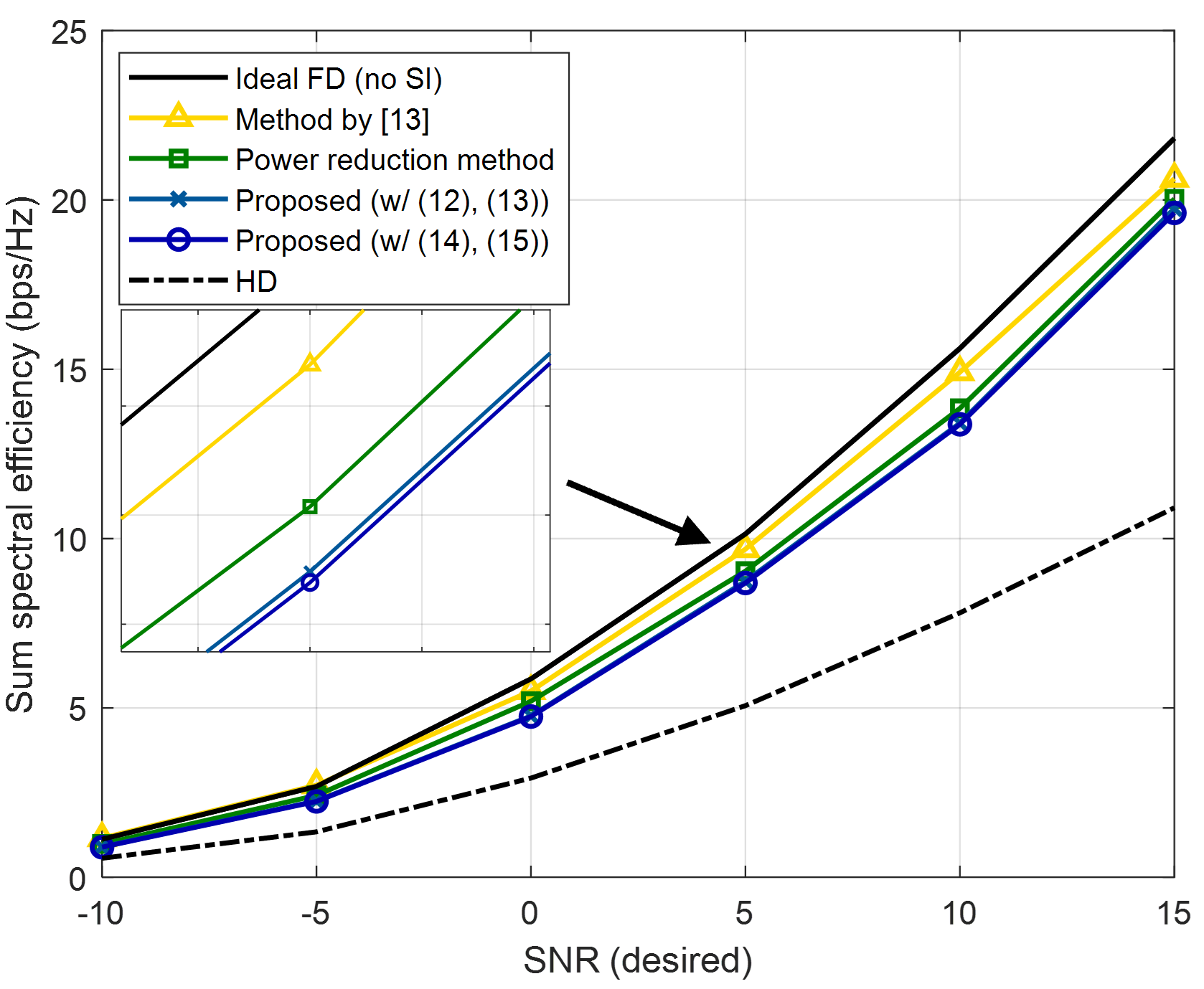}}}
	\caption{Sum spectral efficiency as a function of SNR ($\text{SNR}_{ij}=\text{SNR}_{ki}$) for various beamforming methods.}
	\label{Fig.result_1}	
\end{figure}

For numerical evaluations, we consider $N_t^{(m)}=N_r^{(n)}=16\times4=64$ (uniform planar array),  $L_t^{(m)}=L_r^{(n)}=2$, and $N_s^{\left(mn\right)}=2$. We use DFT codebooks for $\tmatrix{\mathcal{F}}{m}{RF}{}$and $\tmatrix{\mathcal{W}}{n}{RF}{}$, where the node index $(m,n)\in \{(i,j),(k,i)\}$. The total number of $U=128$ subcarriers were considered. For the LNAs and the ADCs, we assume $P^{max}_\text{LNA}=-10$~dBm, $P^{max}_\text{ADC}=-25.98$~dBm. The input power threshold of the $B$ bit ADC was determined by its dynamic range $6.021\cdot B +1.763$~dB, noise floor NF \SI{-90}{dBm}, and an extra \SI{10}{dB} power margin to account for the power-to-average-ratio (PAPR) of the OFDM system, $P^{max}_\text{ADC}=\text{NF}+6.021\cdot B +1.763-10$~dBm.
The transmit power of the FD node is set to $P_{\text{tx}}^{\left(i\right)}=\SI{40}{dBm}$ and the vertical separation of T/Rx arrays to $r^{\left(i\right)}=\SI{0.1}{m}$


\begin{figure}[t]
	\centerline{\resizebox{0.95\columnwidth}{!}{\includegraphics{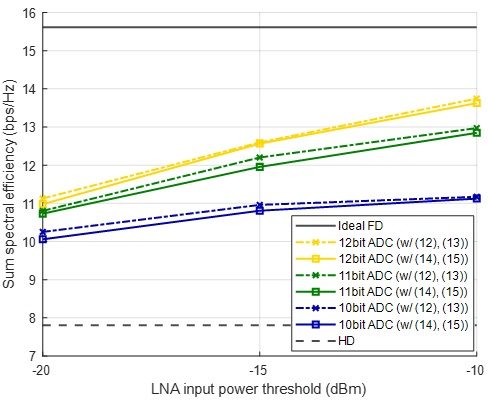}}}
	\caption{{Sum spectral efficiency of proposed methods at $\SI{10}{dB}$ SNR ($\text{SNR}_{ij}=\text{SNR}_{ki}$) under various configurations of Rx components.}}
	\label{Fig.result_2}	
\end{figure}

Fig.~\ref{Fig.result_1} illustrates the sum spectral efficiency of transmit/receive link according to the varying mmWave FD design methods. We observe that our proposed designs shows a performance comparable to the ideal FD and other benchmarks, while requiring less than 63\% of the transmit link beam measurements and 82\% of the total beam measurements compared to those of the benchmarks. Also, the proposed method with \eqref{eq:LNA_const_C4} and \eqref{eq:ADC_const_C4} constraints used only 0.29\% amount of the CPU run time when compared to the method proposed by \cite{ipr2021TWC} to achieve a comparable performance.

Fig.~\ref{Fig.result_2} illustrates the effect of varying $P^{max}_\text{LNA}$ and $P^{max}_\text{ADC}$. It could be intuitively understood that strict LNA and ADC constraints result in allowlists with smaller sizes. Thus, although we also benefit from the reduced number of required beam measurements, sacrifices are being made at the transmit link by excluding the beam candidates that potentially could have provided a strong link to the desired receiver (node $j$). 
Furthermore, the narrow performance gap between the proposed method with \eqref{eq:LNA_const_C3} and \eqref{eq:ADC_const_C3} constraints and the method with \eqref{eq:LNA_const_C4} and \eqref{eq:ADC_const_C4} constraints implies that the sufficient conditions derived in Theorem~\ref{theorem} are practically very tight. Since Theorem~\ref{theorem} allows us to significantly reduce the computational burden imposed by \eqref{eq:LNA_const_C3} and \eqref{eq:ADC_const_C3}, the proposed method with \eqref{eq:LNA_const_C4} and \eqref{eq:ADC_const_C4} constraints would be the preferred candidate for cost-efficient mmWave FD system designs.

\section*{Acknowledgment}
This research was in part supported by Networks Business, Samsung Electronics and Institute of Information \& communications
Technology Promotion (IITP) grant
funded by the Korea government (MSIT) (No. 2021-0-00486, 2021-0-02208).


\section{Conclusion}
\label{sec.conclusion}
In this paper, we have presented a cost-efficient design of the mmWave FD systems. The key design is to dynamically reduce the RF beam codebook in a computationally efficient manner, so that it is comprised of the RF beams that collectively prevent the Rx saturation from the SI. Then, the residual SI can be completely removed with the digital SIC, allowing the digital beamformer to concentrate on its desired channel, free of the SI. To reduce the computation required for the proposed method, we introduced two sufficient conditions that prevent the Rx side saturations, which are sufficiently tight. Through performance evaluations conducted in practical mmWave FD scenarios, realistically modeled via 3D ray-tracing, we demonstrated that the proposed design achieves comparable performance with the ideal FD and other benchmarks with significantly reduced costs.


 




\bibliographystyle{IEEEtran}
\vspace{-6pt}
\bibliography{bibtex_YearSorted}

\begin{thebibliography}{10}
\providecommand{\url}[1]{#1}
\csname url@samestyle\endcsname
\providecommand{\newblock}{\relax}
\providecommand{\bibinfo}[2]{#2}
\providecommand{\BIBentrySTDinterwordspacing}{\spaceskip=0pt\relax}
\providecommand{\BIBentryALTinterwordstretchfactor}{4}
\providecommand{\BIBentryALTinterwordspacing}{\spaceskip=\fontdimen2\font plus
\BIBentryALTinterwordstretchfactor\fontdimen3\font minus
  \fontdimen4\font\relax}
\providecommand{\BIBforeignlanguage}[2]{{%
\expandafter\ifx\csname l@#1\endcsname\relax
\typeout{** WARNING: IEEEtran.bst: No hyphenation pattern has been}%
\typeout{** loaded for the language `#1'. Using the pattern for}%
\typeout{** the default language instead.}%
\else
\language=\csname l@#1\endcsname
\fi
#2}}
\providecommand{\BIBdecl}{\relax}
\BIBdecl

\bibitem{Samsung2014mmWave}
W.~Roh, J.-Y. Seol, J.~Park, B.~Lee, J.~Lee, Y.~Kim, J.~Cho, K.~Cheun, and
  F.~Aryanfar, ``Millimeter-wave beamforming as an enabling technology for {5G}
  cellular communications: Theoretical feasibility and prototype results,''
  \emph{{IEEE} Commun. Mag.}, vol.~52, no.~2, pp. 106--113, Feb. 2014.

\bibitem{RHeath2014spatialySparse}
O.~E. Ayach, S.~Rajagopal, S.~Abu-Surra, Z.~Pi, and R.~W. Heath~Jr.,
  ``Spatially sparse precoding in millimeter wave {MIMO} systems,''
  \emph{{IEEE} Trans. Wireless Commun.}, vol.~13, no.~3, pp. 1499--1513, Mar.
  2014.

\bibitem{fodor2021guest}
G.~Fodor, C.-B. Chae, R.~Wichman, A.~Sabharwal, H.~A. Suraweera, R.~Rao, and
  H.~Alves, ``Guest editorial: Full duplex communications theory,
  standardization, and practice,'' \emph{{IEEE} Wireless Commun.}, vol.~28,
  no.~1, pp. 10--11, Feb. 2021.

\bibitem{Sachin2013Full}
D.~Bharadia, E.~McMilin, and S.~Katti, ``Full duplex radios,'' in \emph{Proc.
  ACM SIGCOMM}, vol.~43, no.~4, New York, NY, USA, 2013, p. 375–386.

\bibitem{kwack}
J.~W. Kwak, M.~S. Sim, I.-W. Kang, J.~S. Park, J.~Park, and C.-B. Chae, ``A
  comparative study of analog/digital self-interference cancellation for full
  duplex radios,'' in \emph{Proc. Asilomar Conf. on Signal, Syst. and Comput.},
  Aug. 2019, pp. 375--386.

\bibitem{prototype}
M.~Chung, M.~S. Sim, J.~Kim, D.~K. Kim, and C.-B. Chae, ``Prototyping real-time
  full duplex radios,'' \emph{{IEEE} Commun. Mag.}, vol.~53, no.~9, pp. 56--63,
  Sep. 2015.

\bibitem{kim2022performance}
S.-M. Kim, Y.-G. Lim, L.~Dai, and C.-B. Chae, ``Performance analysis of
  self-interference cancellation in full-duplex massive {MIMO} systems:
  Subtraction versus spatial suppression,'' \emph{to appear in {IEEE} Trans.
  Wireless Commun.}, 2022.

\bibitem{ipr2021magazine}
I.~P. Roberts, J.~G. Andrews, H.~B. Jain, and S.~Vishwanath, ``Millimeter-wave
  full duplex radios: New challenges and techniques,'' \emph{{IEEE} Wireless
  Commun.}, vol.~28, no.~1, pp. 36--43, Feb. 2021.

\bibitem{ZXiao2017FDmmWave}
Z.~Xiao, P.~Xia, and X.-G. Xia, ``Full-duplex millimeter-wave communication,''
  \emph{{IEEE} Wireless Commun.}, vol.~24, no.~6, pp. 136--143, Dec. 2017.

\bibitem{Hanzo2019TVT}
K.~Satyanarayana, M.~El-Hajjar, P.-H. Kuo, A.~Mourad, and L.~Hanzo, ``Hybrid
  beamforming design for full-duplex millimeter wave communication,''
  \emph{{IEEE} Trans. Veh. Technol.}, vol.~68, no.~2, pp. 1394--1404, Feb.
  2019.

\bibitem{ipr2020bfc}
I.~P. Roberts and S.~Vishwanath, ``Beamforming cancellation design for
  millimeter-wave full-duplex,'' in \emph{Proc. IEEE Glob. Commun. Conf.
  (GLOBECOM)}, 2019, pp. 1--6.

\bibitem{ipr2020freq}
I.~P. Roberts, H.~B. Jain, and S.~Vishwanath, ``Frequency-selective beamforming
  cancellation design for millimeter-wave full-duplex,'' in \emph{Proc. IEEE
  Int. Conf. on Commun. (ICC)}, 2020, pp. 1--6.

\bibitem{ipr2021TWC}
I.~P. Roberts, J.~G. Andrews, and S.~Vishwanath, ``Hybrid beamforming for
  millimeter wave full-duplex under limited receive dynamic range,''
  \emph{{IEEE} Trans. Wireless Commun.}, vol.~20, no.~12, pp. 7758--7772, Dec.
  2021.

\bibitem{DSIC}
E.~Ahmed, A.~M. Eltawil, and A.~Sabharwal, ``Self-interference cancellation
  with nonlinear distortion suppression for full-duplex systems,'' in
  \emph{Proc. Asilomar Conf. on Signal, Syst. and Comput.}, 2013, pp.
  1199--1203.

\bibitem{CVXtool}
M.~Grant and S.~Boyd, ``{CVX}: Software for disciplined convex programming,
  version 2.1,'' Mar. 2014 [Online]. http://cvxr.com/cvx/.

\bibitem{comboGeneral}
J.~Wood, ``High performance tools for combinatorics and computational
  mathematics,'' Mar. 2022 [Online].
  //https://cran.r-project.org/web/packages/RcppAlgos/RcppAlgos.pdf.

\bibitem{Suk2020FDIAB}
G.~Y. Suk, S.-M. Kim, J.~Kwak, S.~Hur, E.~Kim, and C.-B. Chae, ``Full-duplex
  integrated access and backhaul for {5G NR}: Analyses and prototype
  measurements,'' \emph{{IEEE} Wireless Commun. (early access)}, 2022.

\end{thebibliography}


\end{document}